\documentclass[12pt]{article}
\usepackage{hyperref}
\hypersetup{
colorlinks=true,
linkcolor=black,
citecolor=black,
urlcolor=cyan,
filecolor=black,
pdfborder={0 0 0},
}

\usepackage{amssymb}
\usepackage{amsfonts}
\usepackage{mathrsfs}
\usepackage{latexsym}
\usepackage{amsmath}
\usepackage{graphics}
\usepackage{graphicx}
\usepackage{sectsty}
\usepackage{setspace}
\usepackage{multirow}

\usepackage[title]{appendix}

\usepackage{tikz}
\usetikzlibrary{matrix,arrows,decorations.pathmorphing}
\usetikzlibrary{positioning}
\vspace{.5in}

\sectionfont{\normalsize\bf}
\subsectionfont{\rm\normalsize}
\def\f{{\mathfrak f}}
\def\be{\begin{equation}}
\def\ee{\end{equation}}
\def\bd{\left|\begin{matrix}}
\def\ed{\end{matrix}\right|}
\input amssym.def
\input amssym
\def\Cop{{\Bbb C}}

\def\O{{\cal O}}
\def\A{{\cal A}}
\def\R{{\cal R}}
\def\h{{\frak h}}
\def\half{{\scriptstyle {1\over 2}}}

\usepackage{cite}

\numberwithin{equation}{section}

\begin{document}

\thispagestyle{empty}
\begin{flushright}
\end{flushright}
\baselineskip=16pt
\vspace{.5in}
{
\begin{center}
{\bf Off-Shell CHY Amplitudes and Feynman Graphs}
\end{center}}
\vskip 1.1cm
\begin{center}
{Louise Dolan}
\vskip5pt

\centerline{\em Department of Physics}
\centerline{\em University of North Carolina, Chapel Hill, NC 27599} 
\bigskip
\bigskip        
{Peter Goddard}
\vskip5pt

\centerline{\em School of Natural Sciences, Institute for Advanced Study}
\centerline{\em Princeton, NJ 08540, USA}
\bigskip
\bigskip
\bigskip
\bigskip
\end{center}

\abstract{\noindent 
A polynomial form is established for the off-shell CHY scattering equations 
proposed by Lam and Yao. Re-expressing this in terms of independent Mandelstam 
invariants provides a new expression for the polynomial scattering equations, 
immediately valid off shell, which makes it evident that they yield the 
off-shell amplitudes given by massless $\phi^3$ Feynman graphs. A CHY 
expression for individual Feynman graphs, valid even off shell, is established through a recurrence 
relation.

\bigskip

\setlength{\parindent}{0pt}
\setlength{\parskip}{6pt}

\setstretch{1.05}
\vfill\eject
\vskip50pt
\section{Introduction}
\label{Introduction}

 Cachazo, He and Yuan (CHY) \cite{CHY0, CHY1, CHY2} have shown that the scattering equations, originally introduced by Fairlie and Roberts \cite{FR}, describe the kinematics of massless particles in an arbitrary space-time dimension, $D$, by proposing remarkable formulae for tree amplitudes, which have been proved for $\phi^3$ theory and for gauge theory \cite{DG1}. In a subsequent paper \cite{DG2}, we showed how to re-express these in terms of a polynomial form for the scattering equations. Here we show how to write this polynomial form so that it is valid when the external particles are off shell, by expressing it in terms of the independent Mandelstam variables associated with a particular cyclic ordering of the external particles, rather than the  set of Mandelstam variables for all possible channels and orderings used in \cite{DG2}. 
 A particular use for these new off-shell polynomial scattering equations is in establishing recurrence relations and we apply them to establish CHY formulae for individual off-shell Feynman diagrams. These formulae for individual diagrams may help illuminate the underlying structure of the CHY formalism. 
 
Consider $N$ massless particles, labelled by $a\in A=\{1,2,\ldots,N\}$,  with momenta, $k_a, a\in A,$ which are on-shell,  $k^2_a=0$, and satisfy momentum conservation,  $\sum_{a\in A}k_a=0$. Introduce
 a variable $z_a\in\Cop$ for each $a\in A$. Then the {\it scattering equations} are the $N$ equations $f_a(z,k)=0, a\in A,$ where
\be
  f_a(z,k)=\sum_{b\in A\atop b\ne a}{k_a\cdot k_b\over z_a-z_b}.\label{SE}
\ee
This system of $N$ equations is M\"obius invariant, and consequently the functions $f_a$ satisfy three identities,
\be 
\sum_{a\in A} f_a(z,k)=0;\qquad\sum_{a\in A}z_a f_a(z,k)=0;\qquad\sum_{a\in A}z_a^2 f_a(z,k)=0,\label{ids}
\ee
so that only $N-3$ are linearly independent. They are equivalent \cite{DG2} to the $N-3$  {\it polynomial scattering equations} $\hat h_m(z,k)=0, $ where 
\be
\hat h_m(z,k)=\sum_{S\subset A\atop |S|=m}k_S^2z_S, \qquad 2\leq m\leq N-2,\label{PSE}
\ee
 the sum is over all $N!/m!(N-m)!$ subsets $S\subset A$ with $m$ elements, and 
\be
k_S=\sum_{b\in S\atop }k_b,\qquad z_S=\prod_{a\in S}z_a, \qquad S\subset A. \label{kS}
\ee
We shall refer to the $\hat h_m(z,k), 2\leq m\leq N-2,$ as the {\it  scattering polynomials}.
In terms of these, the CHY formula for the sum of planar tree amplitudes in $\phi^3$ is
\be
\A_N=\oint\prod_{m=2 }^{N-2}{1\over \hat h_m(z,k)}\prod_{a<b} (z_a-z_b)\prod_{a\in A}{dz_a\over (z_a-z_{a+1})^2}\bigg/d\omega\,,\label{amp1}
\ee
where $d\omega=dz_idz_jdz_k/(z_i-z_j)(z_i-z_k)(z_j-z_k)$ is the M\"obius invariant measure corresponding to fixing $z_i,z_j,z_k$, and the integral encircles the zeros of $\hat h_m(z,k), 2\leq m\leq N-2.$
Taking $z_1\rightarrow\infty, z_2=1, z_N= 0$, we have
\be
\A_N=\oint{1\over z_{N-1}}\prod_{m=1 }^{N-3}{1\over \mathring{h}_m
(z,k)}\prod_{2\leq a<b\leq N-1} (z_a-z_b)\prod_{a=2}^{N-2}{z_adz_{a+1}\over (z_a-z_{a+1})^2}.\label{amp2}
\ee
where
\be
\mathring{h}_m(z,k)=\lim_{z_1\rightarrow\infty}{\hat h_{m+1}/ z_1}\,,\label{defh}
\ee
and the integration contour now encircles the zeros of  $\mathring{h}_m(z,k), 1\leq m\leq N-3,$ and amounts to a sum over those zeros, of which there are $(N-3)!$. Using the polynomial form of the equation is less singular and more stable for numerical calculations in particular. 
Note that the polynomials $\hat h_m$ are invariant under simultaneous permutations of the momenta, $k_a$, and the variables, $z_a$, whereas the integral depends on the specific choice of the ordering of the momenta, $k_1,k_2,\ldots, k_N,$ and to get the complete tree amplitude one has to sum over the possible orderings. 

The definition (\ref{PSE}) of $\hat h_m$ involves  the $2^N-2N-2$ Mandelstam invariants but, given momentum conservation and the massless condition, just $\half N(N-3)$ of these can be taken to be independent, provided that $N-1\leq D$. We can re-express $\hat h_m$ in terms of an independent set of invariants but not while maintaining its manifest permutation symmetry. We can pick an independent set of momenta as follows. Given the specified order of the momenta, $k_1,k_2,\ldots,k_N,$  consider consecutive sets, 
\be\label{IJ}
[I,J]	= \{a: I\leq a\leq J\}, \quad 1\leq I\leq J<N,
\ee
 There are $\half N(N-1)$ independent invariants, $k^2_{[I,J]}$, if we do not impose the on-shell condition, $k_a^2=0, 1\leq a\leq N$. Imposing that condition leaves the $\half N(N-3)$ invariants, $k_{[I,J]}^2, \,1\leq I< J<N$. In section \ref{OSE}, using  $k_a^2=0$, we show that $\hat h_m$ can be rewritten as
\be
\tilde h_m= \sum_{1\leq I<J<N\atop (I,J)\ne (1,N-1)}k^2_{[I,J]}\,
(z_{I}-z_{I-1}) (z_{J}-z_{J+1})\Pi_{[I,J]^o}^{m-2}
\qquad 2\leq m\leq N-2\,,\label{ospf}
\ee
where $I-1$ is identified with $N$ when $I=1$,  $[I,J]^o$ is the complement of $\{I-1,I,J,J+1\}$ in $A$, and $\Pi_V^n$ 
is the symmetric function,
\be
\Pi_V^n=\sum_{i_1<i_2<\cdots<i_n\atop i_a\in V}z_{i_1}z_{i_2}\cdots z_{i_n}
\ee
where $V\subset A$ and $n\leq |V|$.
Thus, $\Pi_V^0=1$ and $\Pi_V^{|V|}=\prod_{i\in V}z_i$.

We use (\ref{ospf}) to define $\tilde h_m$ as the off-shell polynomial form of the scattering equations, using (\ref{PSE}) only on shell. Note that $\tilde h_m=\hat h_m$ in general only if $k_a^2=0, a\in A$. Both  $\tilde h_m, \,2\leq m\leq N-2,$ and the value of the integral expression (\ref{amp1}) for $\A^N$ only involve the consecutive invariants, $k^2_{[I,J]}$, $1\leq I<J<N$. This implies that, if we replace  $\hat h_m$ with $\tilde h_m$ in
(\ref{amp1}), the integral
\be
\A_N=\oint\prod_{m=2 }^{N-2}{1\over \tilde h_m(z,k)}\prod_{a<b} (z_a-z_b)\prod_{a\in A}{dz_a\over (z_a-z_{a+1})^2}\bigg/d\omega\,,\label{amp3}
\ee
gives the sum of planar tree diagrams for massless $\phi^3$ theory, even when the external momenta are off shell. 

To show this, note that if the momenta, $k_a, 1\leq a\leq N,$ satisfy momentum conservation, but are not necessarily massless, take another set of momenta,  $\ell_a, 1\leq a\leq N,$ satisfying momentum conservation, and such that  $\ell^2_{[I,J]}=k^2_{[I,J]}$, $1\leq I<J<N$, and $\ell_a^2=0, 1\leq a\leq N$. Then $\tilde h_m$ as defined by (\ref{ospf}) is the same for $\{k_a\}$ and for $\{\ell_a\}$. For $\{\ell_a\}$, but not $\{k_a\}$, 
$\tilde h_m$ as defined by (\ref{ospf}) is the same as 
$\hat h_m$ as defined by (\ref{PSE}). Then, for $\hat h_m$ with momenta $\{\ell_a\}$, (\ref{amp1}) gives a sum of planar tree diagrams, each of which is a product of propagators $1/\ell_U^2$, where $U$ is a consecutive subset of $A$. Hence, this equals the corresponding product of propagators   $1/k_U^2$. It follows that  (\ref{amp3}) gives the sum of planar diagrams with the off-shell momenta. 

In section \ref{OSE}, we also show that the off-shell scattering equations derived in this way, $\tilde h_m=0$, with $\tilde h_m$ as in (\ref{ospf}), are the polynomial form of the off-shell scattering equations proposed by  Lam and Yao \cite{LY}.

In section \ref{IFD}, as an application of these off-shell scattering equations we establish CHY integrals associated for individual Feynman tree diagrams using an off-shell recurrence relation. Associating as usual the momentum $k_a$ and complex variable $z_a$ with the $a$-th external leg of an individual Feynman diagram, $\Delta$, we define the corresponding integrand by first associating a cross-ratio with each internal line or propagator of the diagram, in the way originally introduced by Koba and Nielsen \cite{KN}. Thus, 
\begin{center}
\begin{tikzpicture}[scale=0.5]
\path (0,0) coordinate (X);
\fill[blue!20!white]  (X) circle (1.5);
\draw [thick]  (X) circle (1.5);
\path (X) ++ (0:1.5) coordinate (X1);
\path (X) ++ (0:3.5) coordinate (Y1);
\draw [thick] (X1) -- (Y1);
\path (X) ++ (70:1.5) coordinate (X2);
\path (X) ++ (70:3.5) coordinate (Y2);
\node [above] at (Y2) {$z_{I-1}$};
\draw [thick] (X2) -- (Y2);
\path (X) ++ (290:1.5) coordinate (X3);
\path (X) ++ (290:3.5) coordinate (Y3);
\node [below] at (Y3) {$z_{J+1}$};
\draw [thick] (X3) -- (Y3);
\path (X) ++ (180:1.5) coordinate (X5);
\path (X) ++ (180:3.5) coordinate (Y5);
\draw [thick] (X5) -- (Y5);
\node [left] at (Y5) {$z_{1}$};
\path (X) ++ (150:1.5) coordinate (X6);
\path (X) ++ (150:3.5) coordinate (Y6);
\draw [thick] (X6) -- (Y6);
\node [left] at (Y6) {$z_{2}$};
\path (X) ++ (210:1.5) coordinate (X7);
\path (X) ++ (210:3.5) coordinate (Y7);
\draw [thick] (X7) -- (Y7);
\node [left] at (Y7) {$z_{N}$};
\path (X) ++ (260:3) coordinate (X8);
\node [gray] at (X8) {$\bullet$} ;
\path (X) ++ (240:3) coordinate (X9);
\node [gray] at (X9) {$\bullet$} ;
\path (X) ++ (100:3) coordinate (X10);
\node [gray] at (X10) {$\bullet$} ;
\path (X) ++ (120:3) coordinate (X11);
\node [gray] at (X11) {$\bullet$} ;
\path (5,0) coordinate (X);
\fill[blue!20!white]  (X) circle (1.5);
\draw [thick]  (X) circle (1.5);
\path (X) ++ (110:1.5) coordinate (X1);
\path (X) ++ (110:3.5) coordinate (Y1);
\draw [thick] (X1) -- (Y1);
\node [above] at (Y1) {$z_{I}$};
\path (X) ++ (70:1.5) coordinate (X2);
\path (X) ++ (70:3.5) coordinate (Y2);
\node [above] at (Y2) {$z_{I+1}$};
\draw [thick] (X2) -- (Y2);
\path (X) ++ (290:1.5) coordinate (X3);
\path (X) ++ (290:3.5) coordinate (Y3);
\node [below] at (Y3) {$z_{J-1}$};
\draw [thick] (X3) -- (Y3);
\path (X) ++ (250:1.5) coordinate (X4);
\path (X) ++ (250:3.5) coordinate (Y4);
\draw [thick] (X4) -- (Y4);
\node [below] at (Y4) {$z_{J}$};
\path (X) ++ (0:3) coordinate (X8);
\node [gray] at (X8) {$\bullet$} ;
\path (X) ++ (20:3) coordinate (X9);
\node [gray] at (X9) {$\bullet$} ;
\path (X) ++ (340:3) coordinate (X10);
\node [gray] at (X10) {$\bullet$} ;
\node at (12,0) {Fig. 1};
\node at (-8,0) {\phantom{Fig. 1}};
\end{tikzpicture}
\end{center}
\be\label{uIJ}
u_{I,J}={(z_I-z_J)(z_{I-1}-z_{J+1})\over (z_I-z_{J+1})(z_{I-1}-z_J)}\,,
\ee
and taking $u_\Delta$ to be the product of such cross-ratios for the $N-3$ propagators of the diagram. 
The corresponding integral is then 
\begin{align}\label{IDelta}
\A_N^\Delta&=\oint  {1\over u_\Delta}
\prod_{1\leq a<b\leq N\atop d(a,b)>2}(z_a-z_b)
\prod_{a=1}^Ndz_a\prod_{m=2}^{N-2}{1\over\tilde h_m}\Bigg/
\;d\omega\,,\qquad N>4\,,
\end{align}
where the contour encircles the zeros of $\tilde h_m$. [See (\ref{IDelta2}) for the case $N=4$, and (\ref{amp4}) for the definition of $d(a,b)$.] Using an off-shell recurrence relation, we show that $\A_N^\Delta$ is the product of factors $1/k^2_{[I,J]}$ for the $N-3$ propagators, and so equals the value of the Feynman diagram for 
 $\phi^3$ field theory. [For other work on associating CHY expressions to individual Feynman diagrams see \cite{ABHY, CKW, GHZ, BBBD, G}.]

\section{Off-Shell Scattering Equations}
\label{OSE}
To rewrite $\hat h_m$, as defined by  (\ref{PSE}), 
in terms of independent invariants we need to write 
each $k_S^2$ in terms of  $k_{[I,J]}^2, \,1\leq I\leq J<N$. First note that, for any three vectors $K_1,K_2,K_3,$
\be
(K_1+K_3)^2=(K_1+K_2+K_3)^2-(K_1+K_2)^2-(K_2+K_3)^2+K_1^2+K_2^2+K_3^2\,.
\ee
From this it follows by induction that, for vectors $K_i, 1\leq i\leq 2n-1$, 
\begin{align} \label{kid}
(K_1+K_3+\dots +K_{2n-1})^2 &= K_{1,2n-1}^2-K_{1,2n-2}^2-K_{2,2n-1}^2
+ K_{1,2n-3}^2+ K_{2,2n-2}^2 + K_{3,2n-1}^2
\cr
&\qquad\qquad+ \cdots + K_1^2+K_2^2 +K_3^2+\dots+ K_{2n-1}^2\nonumber\\
&=\sum_{1\leq i\leq j\leq 2n-1}(-1)^{i-j}K^2_{i,j}\,,
\end{align}
where $K_{i,j}=K_i+K_{i+1}+\ldots+K_j$.

If $S\subset A$ and $\overline S$ denotes the complement of $S$ in $A$, then $k_{\overline S}^2=k_S^2$. Just one of $S$ and $\overline S$ does not contain $N$. So, writing $A^\ast=\{1,2,\ldots,N-1\},$ we can rewrite (\ref{PSE}) as
\be
\hat h_m(z,k)= \sum_{S\subset A^\ast\atop |S|=m}k_S^2z_S+
\sum_{S\subset A^\ast\atop |\overline S|=m}k_S^2z_{\overline S}
\,,\qquad 2\leq m\leq N-2\,.\label{PSE2}
\ee
Now, we can write a given $S\subset A^\ast$ as the union of consecutive subsets $S_r, r= 1,3,\ldots,2n-1,$ of $A^\ast$, and denote the indices between $S_{2r-1}$ and $S_{2r+1}$ by $S_{2r}$, $1\leq r\leq n-1$, which is nonempty. Taking $K_r=k_{S_r}$,
\be
k_S^2= \sum_{1\leq i\leq j\leq 2n-1}(-1)^{i-j}k^2_{S_{i,j}}\,,\label{kS2}
\ee
where $S_{i,j}=S_i\cup S_{i+1}\cup\ldots S_j$ is also consecutive. Substituting (\ref{kS2}) into (\ref{PSE2}), if $1\leq I<J<N$, and the consecutive subset $[I,J]\subset A^\ast$ occurs as $S_{i,j}$, with $i,j$ both odd, $z_S$ contains a factor $z_Iz_J$ and $z_{\overline S}$ contains a factor $z_{I-1}z_{J+1}$, while if $i,j$ are both even, $z_S$ contains a factor $z_{I-1}z_{J+1}$, and $z_{\overline S}$ contains a factor $z_{I}z_{J}$; if $i$ is odd and $j$ is even  $z_S$ contains a factor $z_Iz_{J+1}$ and $z_{\overline S}$ contains a factor $z_{I-1}z_{J}$, while if $i$ is even and $j$ is odd it is the other way round. If $i$ and $j$ have different parity, the term has a negative sign. For example, with $N=7, S=\{1,4,5\},$ we have $S_1=\{1\},S_2=\{2,3\},S_3=\{4,5\}$, and $S_{1,2}=\{1,2,3\}=S_{[I,J]}$ with $I=1,J=3$ and $z_1z_4$ is a factor of $z_S$, while $z_7z_3$ is a factor of  $z_{\overline S}$ (where $z_{-1}$ is cyclically identified with $z_7$).

In $\hat h_m$, each of the terms $z_{I}z_{J}, z_{I-1}z_{J+1}, -z_{I-1}z_{J}, -z_{I}z_{J+1}$ occurs multiplied by each of the products of $m-2$
distinct factors taken from $[I,J]^o$, yielding $\hat h_m=\tilde h_m,$ where
\be
\tilde h_m= \sum_{1\leq I<J<N\atop (I,J)\ne (1,N-1)}k^2_{[I,J]}\,
(z_{I}-z_{I-1}) (z_{J}-z_{J+1})\Pi_{[I,J]^o}^{m-2}
\qquad 2\leq m\leq N-2\,,\label{ospf2}
\ee
provided that $k^2_a=0$ for each $a\in A$. Relaxing this condition, we take the vanishing of (\ref{ospf2}) as the definition of the off-shell polynomial scattering equations because, inserted in (\ref{amp3}), they give the correct off-shell tree diagrams, as argued in the Introduction.

For example, 
\begin{align}
N&=4\cr
&\tilde h_2= s_{12}z_{14}z_{23} + s_{23}z_{21}z_{34},\cr
N&=5\cr
&\tilde h_2= s_{12}z_{15}z_{23}+ s_{123}z_{15}z_{34} + s_{23}z_{21}z_{34}+ s_{234}z_{21}z_{45}
+ s_{34} z_{32}z_{45} ,\cr
&\tilde h_3= s_{12}z_{15}z_{23}z_4+ s_{123}z_{15}z_{34}z_2 + s_{23}z_{21}z_{34}z_5+ s_{234}z_{21}z_{45}z_3
+ s_{34} z_{32}z_{45} z_1,
\label{wte}\end{align}
where $z_{ab}=z_a-z_b$ and $s_{a_1\ldots a_m}=k^2_{\{a_1,\ldots, a_m\}}$. The cyclic symmetry of $\tilde h_2,\tilde h_3$ is evident if it is noted that $s_{123}=s_{45}$, etc.

To fix the M\"obius invariance, let $z_1\rightarrow\infty, z_2=1, z_N=0,$ and now write
\begin{align}\label{Mfix}
h_m=\lim_{z_1\rightarrow \infty} \tilde h_{m+1}/z_1&= \sum_{J=2}^{N-2}k^2_{[1,J]}\, (z_{J}-z_{J+1})\Pi_{[1,J]^o}^{m-1}
-  \sum_{J=3}^{N-1}k^2_{[2,J]}\,(z_{J}-z_{J+1})\Pi_{[2,J]^o}^{m-1}\cr
&\qquad+ \sum_{3\leq I<J<N}k^2_{[I,J]}\,(z_{I}-z_{I-1}) (z_{J}-z_{J+1})\Pi_{[I,J]^{\prime o}}^{m-2}\,,
\end{align}
$1\leq m\leq N-3$, where $[I,J]^{\prime o}=[I,J]^o\cap A',$ $A'= \{2,3,\ldots, N\}$. 

For the simplest case, $N=4$, writing $z_3=x$, and $s_{ij}=k_{\{i,j\}}^2$, we only have $h_1$, 
\be
h_1=s_{12}\, (1-x)-s_{23}x\,,
\ee
whereas the `on-shell' polynomial $\mathring{h}_1$, defined by (\ref{defh}), is
\be
\mathring{h}_1=s_{12}+s_{13}x
\ee
Since $s_{12}+s_{13}+s_{23}=k_1^2+k_2^2+k_3^2+k_4^2$, $\mathring{h}_1=h_1$ on shell, and
\be
\oint {dx\over x(1-x)h_1}=-{1\over s_{12}}-{1\over s_{23}}\,,\qquad \oint {dx\over x(1-x) \mathring{h}_1}=-{1\over s_{12}}+{1\over s_{12}+s_{13}}\,,
\ee
agreeing only on shell.

With the definition (\ref{ospf2}) of the off-shell scattering equations, $\tilde h_m=0$, we can reverse our calculation to determine how (\ref{PSE}) must be modified to hold off shell. To do this we must remove the terms involving $k_a^2$ in the expression (\ref{kS2}) of $k_S^2$ in terms of the independent invariants $k_{[I,J]}^2$,
\be
k_S^2= \sum_{1\leq i\leq j\leq 2n-1\atop |S_{i,j}|>1}(-1)^{i-j}k^2_{S_{i,j}} +\sum_{a\in S\atop a-1,a+1\notin S}k_a^2+\sum_{a\notin S\atop a-1,a+1\in S}k_a^2\,.
\label{kS3}
\ee
It follows that if we modify (\ref{PSE}) to read
\be
\tilde h_m=\sum_{S\subset A\atop |S|=m}\sigma_Sz_S, \qquad 2\leq m\leq N-2,\label{PSE3}
\ee
where
\be
\sigma_S=
k_S^2-\sum_{a\in S\atop a-1,a+1\notin S}k_a^2-\sum_{a\notin S\atop a-1,a+1\in S}k_a^2\, ,
\label{sS}
\ee
(\ref{PSE3}) is equivalent to (\ref{ospf2}) as a definition of the off-shell polynomial scattering equations. Note that, if $|S|=1$ or $0$, $\sigma_S$, as defined by (\ref{sS}), equals $0$.

To see how the original form of the scattering equations (\ref{SE}) should be modified in order to work off shell, we reverse the line of argument that was used in the derivation of the polynomial form in \cite{DG2}. Consider
\begin{align}\label{thz}
\tilde h_{(z)}&=
\sum_{m=2}^{N-2} (-z)^{N-m-2} \tilde h_m\cr
&=\sum_{m=0}^{N-2} (-z)^{N-m-2} \sum_{S\subset A\atop |S|=m}z_S\left[
k_{\overline S}^2-\sum_{a\in S\atop a-1,a+1\notin S}k_a^2-\sum_{a\notin S\atop a-1,a+1\in S}k_a^2\right],\qquad\hbox{using }k^2_S=k^2_{\overline S},\nonumber\\
&=\prod_{c\in A}(z_c-z)\sum_{a,b\in A\atop a\ne b}{k_a\cdot k_b\over (z-z_a)(z-z_b)}+\sum_{m=0}^{N-2} (-z)^{N-m-2} \sum_{S\subset A\atop |S|=m}\rho_Sz_S \nonumber
\end{align}
where
\begin{align}
\rho_S&=\sum_{a\notin S}k_a^2-\sum_{a\in S\atop a-1,a+1\notin S}k_a^2-\sum_{a\notin S\atop a-1,a+1\in S}k_a^2\cr
&=-\sum_{ a-1,a+1\notin S}k_a^2+\sum_{a\notin S}k_a^2+\sum_{a\notin S\atop a-1,a+1\notin S}k_a^2-\sum_{a\notin S\atop a-1,a+1\in S}k_a^2\cr
&=-\sum_{ a-1,a+1\notin S}k_a^2+\sum_{a,a+1\notin S}k_a^2+\sum_{a-1,a\notin S}k_a^2\,.\\ \nonumber
\end{align}
\vskip-24pt
Thus
\begin{align}\label{osse}
\tilde h_{(z)}\prod_{c\in A}(z_c-z)^{-1}  &=\sum_{a,b\in A\atop a\ne b}{k_a\cdot k_b\over (z-z_a)(z-z_b)}-\sum_{a\in A}{k_a^2\over (z-z_{a-1})(z-z_{a+1})}\cr 
&\quad+\sum_{a\in A}{k_a^2\over (z-z_{a-1})(z-z_{a})}+\sum_{a\in A}{k_a^2\over (z-z_{a})(z-z_{a+1})}\\ \nonumber
\end{align}
\vskip-24pt
Thus the off-shell polynomial scattering equations are equivalent to the identical vanishing of the right hand side of (\ref{osse}). Since this only has single poles and vanishes for large $z$, this is equivalent to the vanishing of its residue at each $z=z_a$, that is to $\f_a(z,k)=0, a\in A$, where
\be
\f_a(z,k)=\sum_{b\ne a} {k_a\cdot k_b\over z_a-z_b}
- {\half k_{a-1}^2 \over z_a-z_{a-2}}+ {\half\left(k_{a-1}^2+k_{a}^2\right)
\over z_a-z_{a-1}}+ {\half\left(k_a^2+k_{a+1}^2\right)\over z_a-z_{a+1}}
-{\half k_{a+1}^2\over z_a-z_{a+2}}\,.\label{lyf1}
\ee
This is the form of the off-shell scattering equations proposed by Lam and Yao \cite{LY}. It follows from our discussion that they do indeed yield the correct off-shell amplitudes. [For other discussion of off-shell amplitudes see \cite{N} and \cite{Y}.]

The M\"obius invariance of the system of off-shell scattering equations follows from the M\"obius invariance of the condition that (\ref{osse}) vanish identically. It can also be verified directly on (\ref{lyf1}). Reflecting this invariance, $\f_a, a\in A,$ satisfy the same relations (\ref{ids}) that the $f_a, a\in A,$ do.

\section{Individual Feynman Diagrams}
\label{IFD}

\subsection{\sl An Example}\label{Ex}

As an example of the recurrence relation we will prove in subsection \ref{RR}, we consider the $N=6$ Mercedes diagram, $M$, shown in Fig. 2a, where
we take $(z_1,z_2,z_3,z_4,z_5,z_6)=(\infty,1,x,y,z,0)$. Then the integral (\ref{IDelta}) takes the form
\be
\A^M_6= \oint_{h_1^6h_2^6h_3^6} {(1-y)(x-z)ydxdydz\over (x-y)z h_1^6h_2^6h_3^6}\,,
\ee
where $h_1^6,h_2^6,h_3^6$ are the M\"obius fixed scattering polynomials defined by (\ref{Mfix}) with $N=6$, 
\begin{align}
&h^6_1= s_{12}z_{23}-s_{23}z_{34}
+s_{123} z_{34} -s_{234}z_{45} +s_{1234}z_{45} - s_{2345}z_5,\cr
&h_2^6= s_{12}z_{23}(z_4+z_5) - s_{23}z_{34}z_5
- s_{34} z_{23}z_{45} - s_{45}z_{34}z_5 + s_{123}z_{34}(z_2+z_5),
\cr
&\hskip25pt - s_{234}z_{45}z_3 -s_{345} z_{23}z_5 + s_{1234}z_{45}(z_2+z_3) - s_{2345}z_5(z_3+z_4)\cr
&h_3^6= s_{12}z_{23}z_4z_5 - s_{45}z_{34}z_5 + s_{123}z_{34}z_5
-s_{345} z_{23}z_4z_5 +s_{1234}z_{45}z_3
-s_{2345}z_3z_4z_5,
\nonumber\end{align}
and 
the integral is round their common zeros.
 \begin{center}
\begin{tikzpicture}[scale=0.5]
\path (0,0) coordinate (X);
\node at (0,-3) {$M$};
\path (X) ++ (0:2) coordinate (Xa);
\draw [thick] (X)--(Xa);
\path (Xa) ++ (45:2) coordinate (Y3);
\path (Xa) ++ (45:3.5) coordinate (Z3);
\node [blue] at (Z3) {$\scriptstyle k_3$};
\draw [thick] (Xa)--(Y3);
\node [above right] at (Y3) {$\scriptstyle z_3=x$};
\path (Xa) ++ (315:2) coordinate (Y4);
\draw [thick] (Xa)--(Y4);
\path (Xa) ++ (315:3.5) coordinate (Z4);
\node [blue] at (Z4) {$\scriptstyle k_4$};
\node [below right] at (Y4) {$\scriptstyle z_4=y$};
\path (X) ++ (90:2) coordinate (Xb);
\draw [thick] (X)--(Xb);
\path (Xb) ++ (135:2) coordinate (Y1);
\path (Xb) ++ (135:3.5) coordinate (Z1);
\node [blue] at (Z1) {$\scriptstyle k_1$};
\draw [thick] (Xb)--(Y1);
\node [above left] at (Y1) {$\scriptstyle z_1=\infty$};
\path (Xb) ++ (45:2) coordinate (Y2);
\draw [thick] (Xb)--(Y2);
\path (Xb) ++ (45:3.5) coordinate (Z2);
\node [blue] at (Z2) {$\scriptstyle k_2$};
\node [above right] at (Y2) {$\scriptstyle z_2=1$};
\path (X) ++ (180:2) coordinate (Xc);
\draw [thick] (X)--(Xc);
\path (Xc) ++ (225:2) coordinate (Y5);
\path (Xc) ++ (225:3.5) coordinate (Z5);
\node [blue] at (Z5) {$\scriptstyle k_5$};
\draw [thick] (Xc)--(Y5);
\node [below left] at (Y5) {$\scriptstyle z_5=z$};
\path (Xc) ++ (135:2) coordinate (Y6);
\draw [thick] (Xc)--(Y6);
\path (Xc) ++ (135:3.5) coordinate (Z6);
\node [blue] at (Z6) {$\scriptstyle k_6$};
\node [above left] at (Y6) {$\scriptstyle z_6=0$};
\path (15,0) coordinate (X);
\node at (15,-3) {$M'$};
\path (X) ++ (0:2) coordinate (Xa);
\draw [thick] (X)--(Xa);
\path (Xa) ++ (0:1.4) coordinate (Z3);
\node [above, blue] at (Z3) {$\scriptstyle k_3+k_4$};
\node [below] at (Z3) {$\scriptstyle z_4=y$};
\path (X) ++ (90:2) coordinate (Xb);
\draw [thick] (X)--(Xb);
\path (Xb) ++ (135:2) coordinate (Y1);
\path (Xb) ++ (135:3.5) coordinate (Z1);
\node [blue] at (Z1) {$\scriptstyle k_1$};
\draw [thick] (Xb)--(Y1);
\node [above left] at (Y1) {$\scriptstyle z_1=\infty$};
\path (Xb) ++ (45:2) coordinate (Y2);
\draw [thick] (Xb)--(Y2);
\path (Xb) ++ (45:3.5) coordinate (Z2);
\node [blue] at (Z2) {$\scriptstyle k_2$};
\node [above right] at (Y2) {$\scriptstyle z_2=1$};
\path (X) ++ (180:2) coordinate (Xc);
\draw [thick] (X)--(Xc);
\path (Xc) ++ (225:2) coordinate (Y5);
\path (Xc) ++ (225:3.5) coordinate (Z5);
\node [blue] at (Z5) {$\scriptstyle k_5$};
\draw [thick] (Xc)--(Y5);
\node [below left] at (Y5) {$\scriptstyle z_5=z$};
\path (Xc) ++ (135:2) coordinate (Y6);
\draw [thick] (Xc)--(Y6);
\path (Xc) ++ (135:3.5) coordinate (Z6);
\node [blue] at (Z6) {$\scriptstyle k_6$};
\node [above left] at (Y6) {$\scriptstyle z_6=0$};\node at (0,-4.5) {Fig. 2a} ;
\node [above left] at (Y6) {$\scriptstyle z_6=0$};\node at (15,-4.5) {Fig. 2b} ;
\end{tikzpicture}
\end{center}
In order to remove the pole of the integrand at $z=0$, we replace $h^6_2, h^6_3, $ by $h^6_{(z_5)}=h^6_{(z)}, h^6_{(z_6)}=h^6_3,$ 
where $h^6_{(z)}=h^6_3-zh^6_2 +z^2h^6_1$ is the $z_1\rightarrow\infty$ limit of (\ref{thz}).
This introduces a Jacobian factor of $-z$, to give
\be
\A^M_6= -\oint_{h_1^6h_2^6h_3^6} {(1-y)(x-z)ydxdydz\over (x-y) h_1^6h_{(z)}^6h_3^6}\,.
\ee
The solutions of $h_1^6=h_2^6=h_3^6=0$ are solutions of $h_1^6=h_{(z)}^6=h_3^6=0$, while solutions of $h_1^6=h_{(z)}^6=h_3^6=0$ are either solutions of $h_1^6=h_2^6=h_3^6=0$  or have $z=0$. From (\ref {Arel3}), we have
$z=0,$ $h^6_{(z)}=s_{56}xy$ and this is cancelled by factors in the numerator so that we may write
\begin{align}
\A^M_6&=-\oint_{h_1^6h_{(z)}^6h_3^6} {(1-y)(x-z)ydxdydz\over (x-y) h_1^6h_{(z)}^6h_3^6}\cr
&= \oint_{h_1^6(x-y)h_3^6} {(1-y)(y-z)ydxdydz\over (x-y) h_1^6h_{(z)}^6h_3^6}\,,
\end{align}
using the global residue theorem. Now, from (\ref {Arel3}), at $x=y,$
\be\label{hhh}
 \left.h_1^6\right|_{x=y}= \h_1^5\,,\quad
 \left.h_{(z)}^6\right|_{x=y}=(y-z) \h_{(z)}^5 -s_{34}(1-y)(z-y)z\,,\quad
  \left.h_3^6\right|_{x=y}= y\h_2^5\,,
  \ee
where $\h^5_1,\h^5_2$, given explicitly in (\ref{h5}), are the scattering polynomials associated with the Feynman diagram, $M'$, shown in Fig. 2b, $\h^5_{(z)}=\h^5_2-z\h^5_1,$
    and so, performing the $x$ integration,
  \begin{align}
\A_6^M
&=-{1\over s_{34}} \oint_{\h_1^5\h_2^5} {dydz\over z\h_1^5\h_2^5}= -{1\over s_{34}} \A_5^{M'}\,,
\end{align}
where $\A_5^{M'}$ is the amplitude associated with $M'$. It straightforward to show 
\be 
\A_5^{M'}={1\over s_{12}s_{56}}\,, \qquad\hbox{and so}\qquad \A_6^{M}=-{1\over s_{12}s_{34}s_{56}}\,.
\ee
This example illustrates the techniques we shall use to establish a general recurrence relation. (For a further example, with $N=8$, see Appendix \ref{Example}.)  But before doing this we need to analyze the structure of the integrand in the next subsection.  

\subsection{\sl Form of the Integrand}\label{FoI}

The sum of off-shell tree diagrams is given by (\ref{amp3}), which can be rewritten
\be
\A_N=\oint\Phi_N\prod_{d(a,b)>2} (z_a-z_b)\prod_{m=2 }^{N-2}{1\over \tilde h_m}\prod_{a\in A}{dz_a}\bigg/d\omega\,,\qquad N>4\,,\label{amp4}
\ee
where $d(a,b)=\min(|a-b|,N-|a-b|), 1\leq a<b\leq N$, and 
\be\label{defPhi}
\Phi_N=\prod_{d(a,b)=2} (z_a-z_b)\bigg/ \prod_{d(a,b)=1} (z_a-z_b)\,.
\ee
We associate the individual Feynman diagram $\Delta$ with the integral, $\A_N^\Delta$, obtained by replacing $\Phi_N$ by $1/u_\Delta$, both of which are M\"obius invariant,
\begin{align}\label{IDelta2}
\A_N^\Delta&=\oint  {1\over u_\Delta}
\prod_{1\leq a<b\leq N\atop d(a,b)>2}(z_a-z_b) \prod_{m=2}^{N-2}{1\over\tilde h_m}
\prod_{a=1}^Ndz_a\Bigg/
\;d\omega\,,\qquad N>4\,,
\end{align}
where $u_\Delta$ is the product of the cross-ratios (\ref{uIJ}). [In the case $N=4$, the first product in the integrands of (\ref{amp4}) and (\ref{IDelta2}) should be replaced by $(z_1-z_3)^{-1}(z_2-z_4)^{-1}$.]

The product $1/u_\Delta$ has only simple poles in the variables $z_a$, and these only occur at locations of the form $z_a=z_b$ where $|a-b|\ne 2$.
To see this is true, suppose $\Delta$ has $i_1$ vertices with one internal leg (type 1),
$i_2$ vertices with two internal legs (type 2), and $i_3$ vertices with three internal legs (type 3), so that total number of vertices, $V=i_1+i_2+i_3=N-2$. As each internal leg or propagator connects two vertices, the number of propagators, $P=\half(i_1+2i_2+3i_3)=N-3$. Thus
\be
i_3=i_1-2,\qquad i_2= N - 2i_1.
\label{nps}\ee
As in Fig. 1, each propagator corresponds to a consecutive subset $S=[I,J]\subset A$, where $1\leq I<J<N$ (and to its complement, but choose the representative subset so that $N\notin S$. The graph has at least two type 1 vertices and we label the graph so that the legs $1,N$ meet at one of these vertices. Then each propagator corresponds to a consecutive subset of $[2,N-1]$, including $[2,N-1]$ itself, and we denote these subsets by $S_i, 1\leq i\leq N-3.$ They are such that,  if $S_i\cap S_j\ne \varnothing$, either $S_i\subset S_j$, or $S_j\subset S_i$. Thus, the propagator subsets, $S_i$, are partially ordered: given a propagator $S_{i_0}$, all the propagators $S_j\supset S_{i_0}$ can be ordered into an ascending sequence, $S_{i_0}\subset S_{i_1}\subset S_{i_2}\subset \ldots \subset [2,N-1] =S_{i_M}, $ for some $M$, with $S_{i_k}$ meeting $S_{i_{k+1}}$ at a vertex, $0\leq k\leq M-1$.

At a type 1 vertex, external legs, $I,I+1,$ meet to form the propagator $[I,I+1]$. Consider the vertex next above the propagator $[I,J]$ in the ascending sequence. If it is a type 2 vertex, either the external leg $I-1$ meets the propagator $[I,J]$ to form the propagator $[I-1,J]$, or the propagator $[I,J]$ meets the external leg $J+1$ to form the propagator $[I,J+1]$. If it is   a type 3 vertex, the propagator $[I,J]$ meets the propagator $[J+1,K]$ to form the propagator $[I,K]$, for some $I<J<K$.

The propagator $[I,J]$ is associated with a factor in $1/u_\Delta$ given by the cross-ratio,
\be
{1\over u_{I,J}} =
{(z_I-z_{J+1}) (z_{I-1}-z_{J})\over (z_I-z_{J}) (z_{I-1}-z_{J+1})}.
\ee 
The factor  $(z_I-z_J)$ only occurs in the denominators of the cross-ratios associated with 
the potential propagators $[I,J]$ and  $[I+1,J-1]$. Thus   $(z_I-z_J)^{-2}$ cannot occur in the 
denominator unless both  $[I,J]$ and $[I+1,J-1]$ actually are propagators, in which case $I<J-2$. 

Taking $S_{i_0}= [I+1,J-1]$, the next propagator,  $S_{i_1}$, in the ascending sequence from  
$S_{i_0}$ cannot be  $[I,J]$, because  $\{I,J\}$ is not a consecutive set, and thus 
$[I,J]$ does not meet $[I+1,J-1]$ at a vertex. 
It follows that  $[I,J]=S_{i_2}$, with either $S_{i_1}=[I,J-1]$ or  $S_{i_1}=[I+1,J]$. 
The cross-ratios associated with $[I,J-1]$ or with $[I+1,J]$ both contain a factor of  
$(z_I-z_J)$ in the numerator, so that, whether $S_{i_1}=[I,J-1]$ or  $S_{i_1}=[I+1,J]$, the product of cross-ratios for $S_{i_0}, S_{i_1},$ and $S_{i_2}$ only has a simple pole at   $z_I=z_J$. Since this does
not occur as a pole in the cross-ratio associated with any other propagator, 
the integrand itself has at most a simple pole at $z_I=z_J$.

The factor $(z_I-z_{I+2})$ only occurs in the denominator of the cross-ratio associated with 
$[I,I+2]$. If this is a propagator, either $[I,I+1]$ or $[I+1,I+2]$ must be
 in the sequence. The cross-ratios of each of these contains a factor of $(z_I-z_{I+2})$ in the numerator, so the product of the cross-ratios for all propagators does not contain a pole at $z_I=z_{I+2}$. Similar considerations exclude poles at $z_2=z_N$ and $z_1=z_{N-1}$. It follows that the product of cross-ratios contains only simple poles and these are at $z_a=z_b$, for some values of $a,b$ satisfying $|a-b|=1$ or $|a-b|>2$. The latter poles are cancelled in the integrand of (\ref{IDelta}) leaving just $i_1$ simple poles at positions $z_I=z_{I+1}$ 
 corresponding to the type 1 vertices.

\subsection{\sl Recurrence Relation}\label{RR}

We now establish a recurrence relation for $\A^\Delta_N$, which we use to establish that it equals the corresponding Feynman diagram. Consider an $N$-point Feynman diagram, $\Delta$, with possibly off-shell momenta, $k_a$, and associated variables $z_a$, $a\in A$, labeled so that the legs associated with $z_3$ and $z_4$ meet in a vertex. Fix M\"obius invariance by setting $z_1\rightarrow\infty, z_2=1, $ and $z_N=0$, and write $x=z_3, y=z_4$. Further consider the $(N-1)$-point diagram, $\Delta'$, obtained by removing the $z_3,z_4$ external legs of $\Delta$, so that the propagator at which they join becomes an external leg and the external legs of $\Delta'$ are labeled $z_1,z_2,z_4,z_5,\ldots,z_N$, and the associated momenta are $k_1,k_2,k_3+k_4,k_5,\ldots,k_N$. [Note that $(k_3+k_4)^2\ne 0$ in general, even if $k_a^2\ne 0$ for $1\leq a\leq N$, so that an off-shell recurrence relation is needed.]
 
After fixing M\"obius invariance, the amplitude associated with $\Delta$ is given by 
\be\label{AND}
\A_N^\Delta=
\oint_{h_1^Nh_2^N \ldots h_{N-3}^N}F_N^\Delta\,{dxdydz_5\dots dz_{N-1}
\over  h_1^Nh_2^N \ldots h_{N-3}^N}
\ee
where the suffix $h_1^Nh_2^N \ldots h_{N-3}^N$ on the integral indicates that it is taken round the common zeros of $h^N_a,1\leq a\leq N-3$,
\be
F_N^\Delta={1\over u_\Delta}
\prod_{2\leq a<b\leq N\atop d(a,b)>2}(z_a-z_b)\,,\label{FN}
\ee
and  $h^N_m=\lim_{z_1\rightarrow \infty} \tilde h^N_{m+1}/z_1,$ the superscript $N$ indicating that these are the scattering polynomials associated to the $N$-point amplitude, with momenta, $k_1,\ldots,k_N$, and variables, $z_1,\ldots,z_N$ rather than the $(N-1)$-point one, which we shall denote by $\h^{N-1}_m$, with momenta, $k_1,k_2,k_3+k_4,k_5,\ldots,k_N$, and variables, $z_1,z_2,z_4,z_5,\ldots,z_N$.
\begin{center}
\begin{tikzpicture}[scale=0.5]
\path (0,0) coordinate (X);
\fill[blue!20!white]  (X) circle (1.5);
\draw [thick]  (X) circle (1.5);
\node at (0,0) {$\Delta$};
\path (X) ++ (90:1.5) coordinate (X1);
\path (X) ++ (90:3.5) coordinate (Y1);
\path (X) ++ (90:4.5) coordinate (Z1);
\draw [thick] (X1)--(Y1);
\node [above] at (Y1) {$\scriptstyle z_1=\infty$};
\node [blue] at (Z1) {$\scriptstyle k_1$};
\path (X) ++ (45:1.5) coordinate (X2);
\path (X) ++ (45:3.5) coordinate (Y2);
\path (X) ++ (45:5) coordinate (Z2);
\node [blue] at (Z2) {$\scriptstyle k_2$};
\draw [thick] (X2)--(Y2);
\node [above right] at (Y2) {$\scriptstyle z_2=1$};
\path (X) ++ (0:1.5) coordinate (X0);
\path (X) ++ (0:3.5) coordinate (Y0);
\draw [thick] (X0)--(Y0);
\path (Y0) ++ (45:2) coordinate (Y3);
\path (Y0) ++ (45:3.5) coordinate (Z3);
\node [blue] at (Z3) {$\scriptstyle k_3$};
\draw [thick] (Y0)--(Y3);
\node [above right] at (Y3) {$\scriptstyle z_3=x$};
\path (Y0) ++ (315:2) coordinate (Y4);
\draw [thick] (Y0)--(Y4);
\path (Y0) ++ (315:3.5) coordinate (Z4);
\node [blue] at (Z4) {$\scriptstyle k_4$};
\node [below right] at (Y4) {$\scriptstyle z_4=y$};
\path (X) ++ (315:1.5) coordinate (X5);
\path (X) ++ (315:3.5) coordinate (Y5);
\path (X) ++ (315:5) coordinate (Z5);
\node [blue] at (Z5) {$\scriptstyle k_5$};
\draw [thick] (X5)--(Y5);
\node [below right] at (Y5) {$\scriptstyle z_5$};
\path (X) ++ (270:1.5) coordinate (X6);
\path (X) ++ (270:3.5) coordinate (Y6);
\path (X) ++ (270:4.5) coordinate (Z6);
\node [blue] at (Z6) {$\scriptstyle k_6$};
\draw [thick] (X6)--(Y6);
\node [below] at (Y6) {$\scriptstyle z_6$};
\path (X) ++ (180:1.5) coordinate (X7);
\path (X) ++ (180:3.5) coordinate (Y7);
\path (X) ++ (180:6) coordinate (Z7);
\node [blue] at (Z7) {$\scriptstyle k_{N-1}$};
\draw [thick] (X7)--(Y7);
\node [left] at (Y7) {$\scriptstyle z_{N-1}$};
\path (X) ++ (135:1.5) coordinate (X8);
\path (X) ++ (135:3.5) coordinate (Y8);
\path (X) ++ (135:5) coordinate (Z8);
\node [blue] at (Z8) {$\scriptstyle k_N$};
\draw [thick] (X8)--(Y8);
\node [above left] at (Y8) {$\scriptstyle z_{N}=0$};
\path (X) ++ (210:3.5) coordinate (B1);
\path (X) ++ (225:3.5) coordinate (B2);
\path (X) ++ (240:3.5) coordinate (B3);
\node [gray] at (B1) {$\bullet$} ;
\node [gray] at (B2) {$\bullet$} ;
\node [gray] at (B3) {$\bullet$} ;
\path (15,0) coordinate (X);
\fill[blue!20!white]  (X) circle (1.5);
\draw [thick]  (X) circle (1.5);
\node at (X) {$\Delta'$};
\path (X) ++ (90:1.5) coordinate (X1);
\path (X) ++ (90:3.5) coordinate (Y1);
\path (X) ++ (90:4.5) coordinate (Z1);
\draw [thick] (X1)--(Y1);
\node [above] at (Y1) {$\scriptstyle z_1=\infty$};
\node [blue] at (Z1) {$\scriptstyle k_1$};
\path (X) ++ (45:1.5) coordinate (X2);
\path (X) ++ (45:3.5) coordinate (Y2);
\path (X) ++ (45:5) coordinate (Z2);
\node [blue] at (Z2) {$\scriptstyle k_2$};
\draw [thick] (X2)--(Y2);
\node [above right] at (Y2) {$\scriptstyle z_2=1$};
\path (X) ++ (0:1.5) coordinate (X0);
\path (X) ++ (0:3.5) coordinate (Y0);
\draw [thick] (X0)--(Y0);
\path (X) ++ (0:5.2) coordinate (Z0);
\node [right, blue] at (Z0) {$\scriptstyle k_3+k_4$};
\node [right] at (Y0) {$\scriptstyle z_4=y$};
\path (X) ++ (315:1.5) coordinate (X5);
\path (X) ++ (315:3.5) coordinate (Y5);
\path (X) ++ (315:5) coordinate (Z5);
\node [blue] at (Z5) {$\scriptstyle k_5$};
\draw [thick] (X5)--(Y5);
\node [below right] at (Y5) {$\scriptstyle z_5$};
\path (X) ++ (270:1.5) coordinate (X6);
\path (X) ++ (270:3.5) coordinate (Y6);
\path (X) ++ (270:4.5) coordinate (Z6);
\node [blue] at (Z6) {$\scriptstyle k_6$};
\draw [thick] (X6)--(Y6);
\node [below] at (Y6) {$\scriptstyle z_6$};
\path (X) ++ (180:1.5) coordinate (X7);
\path (X) ++ (180:3.5) coordinate (Y7);
\path (X) ++ (180:6) coordinate (Z7);
\node [blue] at (Z7) {$\scriptstyle k_{N-1}$};
\draw [thick] (X7)--(Y7);
\node [left] at (Y7) {$\scriptstyle z_{N-1}$};
\path (X) ++ (135:1.5) coordinate (X8);
\path (X) ++ (135:3.5) coordinate (Y8);
\path (X) ++ (135:5) coordinate (Z8);
\node [blue] at (Z8) {$\scriptstyle k_N$};
\draw [thick] (X8)--(Y8);
\node [above left] at (Y8) {$\scriptstyle z_{N}=0$};
\path (X) ++ (210:3.5) coordinate (B1);
\path (X) ++ (225:3.5) coordinate (B2);
\path (X) ++ (240:3.5) coordinate (B3);
\node [gray] at (B1) {$\bullet$} ;
\node [gray] at (B2) {$\bullet$} ;
\node [gray] at (B3) {$\bullet$} ;
\node at (8,-5.5) {Fig. 3} ;
\end{tikzpicture}
\end{center}
The amplitude associated with $\Delta'$ is given by 
\be
\A_{N-1}^{\Delta'}=
\oint_{\h_1^{N-1}\h_2^{N-1} \ldots \h_{N-4}^{N-1}}F_{N-1}^{\Delta'}\,{dydz_5\dots dz_{N-1}\,.
\over  \h_1^{N -1}\h_2^{N -1}\ldots \h_{N-4}^{N-1}}
\ee
The  extra cross-ratio factor in $F^\Delta_N$ relative to $F_{N-1}^{\Delta'}$ 
corresponds to the propagator factor 
\be
{(x-z_5)(1-y)\over (x-y)(1-z_5)}
\ee
 and  other factors from the product in (\ref{FN}) are $(1-z_5)$, $(x-z_6)$, $ (x-u_a),$ $7\leq a\leq N,$ so that
\be\label{Free}
F^\Delta_N={1-y\over x-y} \prod_{a=5}^{N}(y-z_a) \left[F^{\Delta'}_{N-1} + \O(x-y)\right]\,.
\ee
To establish the recurrence relation, as in subsection \ref{Ex}, it is useful to make use of the function $\tilde h_{(z)}$ introduced in (\ref{thz}), or rather its asymptotic form as 
$z_1\rightarrow\infty$, 
\be
h_{(z)}=\lim_{z_1\rightarrow\infty}\left(\tilde h^N_{(z)}\big/z_1\right)=
\sum_{m=1}^{N-3} (-z)^{N-m-3}h_{m}^N\,.
\ee
Replacing $ h_1^Nh_2^N \ldots h_{N-3}^N$ by $h_1^N h^N_{(z_5)} \ldots h^N_{(z_{N-1})}h^N_{(z_N)}$ in the denominator of (\ref{AND}) introduces a Jacobian factor of 
\be
(-1)^{\half N(N-1)}\prod_{5\leq a<b\leq N}(z_a-z_b)
\ee
 when the integral is evaluated at $ h_1^N=h_2^N = \ldots = h_{N-3}^N=0,$ so that
 \be\label{ADNp}
\A^\Delta_N=(-1)^{\half N(N-1)}
\oint_{h_1^Nh_2^N \ldots h_{N-3}^N}F^\Delta_N\, \prod_{5\leq a<b\leq N}(z_a-z_b)\, {dxdydz_5\dots dz_{N-1}
\over  h_1^N h^N_{(z_5)}h^N_{(z_{6})} \ldots h^N_{(z_N)}}\,.
\ee
$ h_1^N, h^N_{(z_5)},h^N_{(z_{6})} ,\ldots, h^N_{(z_N)}$ all vanish when $h_1^N, h_2^N \ldots h_{N-3}^N$ all vanish. Conversely, if $h_1^N=0$ and $h^N_{(z_a)}=0, 5\leq a\leq N,$ then $h^N_m=0, 1\leq m\leq N-3$, unless $z_a=z_b$ for some $a,b$ with $5\leq a<b\leq N$, but in such a case the contribution to the integrand from such a point is cancelled by the product of such factors $z_a-z_b$ in the integrand. From (\ref{Free}), we see that $F_N^\Delta$ has a pole at $x=y$ and, from subsection \ref{FoI}, we know that its other possible singularities are simple poles at $z_{a-1}=z_{a}$, where $6\leq a\leq N$. However we shall show that such poles are cancelled by a zero in $F_N^\Delta$, and so they do not contribute to the integral. Assuming this to be the case, and using (\ref{Free}),
\begin{align}
\A^\Delta_N&= (-1)^{\half N(N-1)}
\oint_{h_1^N h_{(z_5)}^N h_{(z_6)}^N \ldots h_{(z_N)}^N} \left[F^{\Delta'}_{N-1} + \O(x-y)\right]{1-y\over x-y} \,\prod_{a=5}^{N}(y-z_a)
\cr
&\qquad\qquad\qquad\qquad\times 
\prod_{5\leq a<b\leq N}(z_a-z_b)\,
{dxdydz_5\dots dz_{N-1}
\over  h_1^N h_{(z_5)}^N h_{(z_{6})}^N\ldots h^N_{(z_N)}}\,.
\end{align}
Now, the only pole of the integrand, other than where  $h_1^N =h_{(z_5)}^N =h_{(z_6)}^N= \ldots =h_{(z_N)}^N=0$,
is at $x=y$ and so, using the global residue theorem,  we can replace $h_{(z_5)}^N$ with $x-y$ in the specification of the integration contour,
\begin{align}
\A^\Delta_N&=- (-1)^{\half N(N-1)}
\oint_{h_1^N (x-y)\, h_{(z_6)}^N \ldots h_{(z_N)}^N} F^{\Delta'}_{N-1}{1-y\over x-y} \prod_{a=5}^{N}(y-z_a)  \cr
&\qquad\qquad\qquad\qquad\qquad\times 
\prod_{5\leq a<b\leq N}(z_a-z_b)\,
{dxdydz_5\dots dz_{N-1}
\over  h_1^N h_{(z_5)}^Nh_{(z_{6})}^N \ldots h^N_{(z_N)}}\cr
&=(-1)^{\half N(N-1)}
\oint_{h_1^N  h_{(z_6)}^N h_{(z_{7})}^N\ldots h_{(z_N)}^N} F^{\Delta'}_{N-1}  (1-y) \prod_{a=5}^{N}(y-z_a) 
\cr
&\qquad\qquad\qquad\qquad\qquad\times
\prod_{5\leq a<b\leq N}(z_a-z_b)\,
{dydz_5\dots dz_{N-1}
\over  [h_1^N h_{(z_5)}^Nh_{(z_{6})}^N \ldots h^N_{(z_N)}]_{x=y}}\,,\qquad\quad
\end{align}
on performing the $x$ integration.  Now, using (\ref{A1}) and (\ref{Arel3}),
\be
 \left.h_1^N\right|_{x=y}= \h_{1}^{N-1}\,,\qquad \left.h_{(z_b)}^N\right|_{x=y}= (y-z_b)\h_{(z_b)}^{N-1},\quad 6\leq b\leq N\,,
\ee
\begin{align}
\A^\Delta_N&= (-1)^{\half N(N-1)}
\oint_{h_1^N  h_{(z_6)}^Nh_{(z_{7})}^N \ldots h_{(z_N)}^N} F^{\Delta'}_{N-1}  \left.{1\over  h^N_{(z_5)}}\right|_{x=y} (1-y) (y-z_5) 
\cr
&\qquad\qquad\qquad\qquad\qquad\times
\prod_{5\leq a<b\leq N}(z_a-z_b)\,
{dydz_5\dots dz_{N-1}
\over  \h_1^{N-1} \h_{(z_6)}^{N-1} \h_{(z_{7})}^{N-1}\ldots \h^{N-1}_{(z_N)}}
\end{align}
 We can now replace $\h_1^{N-1} \h_{(z_6)}^{N-1}\h_{(z_{7})}^{N-1} \ldots \h^{N-1}_{(z_N)}$
 with $\h_1^{N-1} \h_{2}^{N-1}\h_{3}^{N-1} \ldots \h^{N-1}_{N-4}$ and an appropriate Jacobian factor,
  \begin{align}\label{ARR}
\A^\Delta_N&= (-1)^{N+1}
\oint_{h_1^N  h_{(z_6)}^N \ldots h_{(z_{N-1})}^Nh_{(z_N)}^N} F^{\Delta'}_{N-1} \left.{1\over h^N_{(z_5)}}\right|_{x=y} (1-y) (y-z_5) 
\cr
&\qquad\qquad\qquad\qquad\qquad\times 
\prod_{b=6}^N(z_5-z_b)\,
{dydz_5\dots dz_{N-1}
\over  \h_1^{N-1} \h_{2}^{N-1} \ldots \h_{N-5}^{N-1}\h^{N-1}_{N-4}}\cr
&= - {1\over s_{34}} 
\oint_{ \h_1^{N-1} \h_{2}^{N-1} \ldots \h_{N-5
}^{N-1}\h^{N-1}_{N-4}} F^\Delta_{N-1} 
{dydz_5\dots dz_{N-1}
\over  \h_1^{N-1} \h_{2}^{N-1} \ldots \h_{N-5}^{N-1}\h^{N-1}_{N-4}}=- {1\over s_{34}} \A^\Delta_{N-1} \,,
 \end{align}
 where we have used (\ref{Arel3})
 \be
 \left.h_{(z_5)}^N\right|_{x= y}= (y-z_5)\h_{(z_5)}^{N-1}
 +s_{34} (1-y)(z_5-y)\prod_{b=6}^{N}(z_b-z_5)\,.
 \ee
 
 To complete the argument, we need to show that there are no contributions to the $\A^\Delta_N$ from poles that $F^\Delta_N$ may have at $z_{a-1}=z_{a}$, for some $6\leq a\leq N$.
To consider the possible contribution to (\ref{ADNp}) from such a pole in $F^\Delta_{N} $, note, as in (\ref{Free}), that
\begin{align}\label{Free2}
F^\Delta_N&={(-1)^{N-a} \over z_{a-1}-z_a} \prod_{b=2\atop b\ne a-1,a}^{N}(z_b-z_a) \left[F^{\Delta''}_{N-1} + \O(z_{a-1}-z_a)\right]\cr
&= (-1)^{N-a} {1\over z_{a-1}-z_a} {1\over s_{a-1,a}}
h^N_{(z_a)}\big|_{z_{a-1}=z_a} \left[F^{\Delta''}_{N-1} + \O(z_{a-1}-z_a)\right]
\end{align}
and $h^N_{(z_a)}$ is one of the zero denominators in $\A^\Delta_{N} $, 
which demonstrates that this contribution to the integral vanishes. 

From (\ref{ARR}), we have the recurrence relation,
\begin{align}
\A_N^\Delta = -{1\over s_{34}} \A_{N-1}^{\Delta'}\,,
\end{align}
which shows by induction that $\A_N^\Delta$ equals the product of factors $1/k^2_{[I,J]}$ for each propagator, {\it i.e.} it equals the value of the Feynman diagram.

Since we know that (\ref{amp4}) gives the sum of $N$-point planar Feynman diagrams, $\Delta$ \cite{DG1, DG2}, it follows that the difference
\be
\R_N= \Phi_N-\sum{1\over u_\Delta}
\ee
must integrate to zero when $\Phi_N$ is replaced by $\R_N$ in (\ref{amp4}). For $N\leq 7$, we can find a convenient expression for $\R_N$ and verify that this integral vanishes. If this could be done for all $N$, this would give a direct proof that the CHY integral (\ref{amp4}) equals the sum of planar Feynman diagrams, as established in  \cite{DG1, DG2}.

\section{Comments}

In non-abelian gauge theory, the expressions we have given would be dressed
by a group factor.
Similar extra factors would also occur if the
scalar field transformed in the adjoint representation of some global internal
symmetry. In the scalar case, to the extent that the complete tree
amplitude is a linear combination of `double-partial amplitudes' with
coefficients given by group factors, and that the `double-partial
amplitudes' are themselves sums of subsets of the
individual ordinary $\phi^3$ planar graphs,
then the off-shell version is given by replacing those with
individual off-shell Feynman graphs.
A key feature of our analysis in this paper is to provide the off-shell formula
for any individual Feynman tree diagram in ordinary $\phi^3$ theory.
To arrive at a consistent picture for individual graphs in the gauge theory
would require suitable M\"obius invariant numerators
with dependence on the polarization and momenta.

\section*{Acknowledgements}

LD thanks the Institute for Advanced Study at Princeton for its hospitality.
LD was partially supported by NSF grant No. PHY-1620311 
and PG was partially supported by NSF grant No. PHY-1314311.

\begin{appendices}
\section{Identities}
\label{Identities}
We consider the off-shell scattering polynomials in variables, $z_1,z_2,\ldots,z_N,$ for an $N$-point amplitude with momenta,  $k_1,k_2,\ldots,k_N.$
Using the formula (\ref{ospf2}) for $\tilde h^N_m$ in the definition (\ref{thz}) of $ \tilde h^N_{(z)}$,
\begin{align}
 \tilde h^N_{(z)}&
=\sum_{m=2}^{N-2}  \tilde h^N_m(-z)^{N-2-m}\cr
&=  \sum_{1\leq I<J<N}k^2_{[I,J]}\,
(z_{I}-z_{I-1}) (z_{J}-z_{J+1})\prod_{a\in [I,J]^o}(z_a-z)
\label{Alch}\end{align}
If we put $z_a=z_{a-1}$, $ \tilde h^N_{(z)}$ is related to the corresponding function $ \tilde \h^{N-1}_{(z)}$ for an $(N-1)$-point amplitude, with momenta, 
$k_1,k_2,\ldots,k_{a-2},k_{a-1}+k_a, k_{a+1}, \ldots, k_N,$ and associated variables, $z_1,z_2,\ldots,z_{a-2},z_a, z_{a+1}, \ldots, z_N,$
by the equation,
\begin{align}
\left.\tilde h^N_{(z)}\right|_{z_{a-1}=z_a}
&=(z_a-z) \tilde{\frak h}^{N-1}_{(z)}
+s_{a-1,a} (z_{a+1}-z_a)(z_{a-2}-z_a)  \prod_{b\ne a-2,a-1,a,a+1} (z_b-z),
\label{Aha}
\end{align} 
where $s_{a-1,a}=k^2_{[a-1,a]}$. From this it follows that
  \begin{align}\label{Arel}
  \left.\tilde h^N_{(z_a)}\right|_{z_{a-1}=z_a}
&=s_{a-1,a}   \prod_{b\ne a-1,a} (z_b-z_a),\cr
\left.\tilde h^N_{(z_b)}\right|_{z_{a-1}=z_a}
&=(z_a-z_b) \tilde{\frak h}^{N-1}_{(z_b)}\,,
\hskip50truemm b\ne a-2,a-1,a,a+1,\cr
\left.\tilde h^N_{(z_{a-2})}\right|_{z_{a-1}=z_a}
&=(z_a-z_{a-2}) \tilde{\frak h}^{N-1}_{(z_{a-2})}
+s_{a-1,a} (z_{a+1}-z_a)(z_{a-2}-z_a)  \prod_{b\ne a-2,a-1,a,a+1} (z_b-z_{a-2}),\cr
\left.\tilde h^N_{(z_{a+1})}\right|_{z_{a-1}=z_a}
&=(z_a-z_{a+1}) \tilde{\frak h}^{N-1}_{(z_{a+1})}
+s_{a-1,a} (z_{a+1}-z_a)(z_{a-2}-z_a)  \prod_{b\ne a-2,a-1,a,a+1} (z_b-z_{a+1}),\cr
  \end{align}
   Above we have used gothic script for the polynomial 
   $\tilde {\h}^{N-1}_m, 2\le m\le N-3$, which differs from $\tilde h^{N-1}_m$,
in that the arguments of  $\tilde {\h}^{N-1}_m$ are
momenta $k_1,k_2,\ldots k_{a-2}, (k_{a-1}+k_a), k_{a+1},\ldots k_N$
with $\sum_{a=1}^Nk_a =0$, whereas $\widetilde h^{N-1}_m$ would have momenta
$k_a, 1\le a\le N-1$, with $\sum_{a=1}^{N-1} k_a =0.$ 
Thus to define $\widetilde {\h}^{N-1}_m$
fully, one  needs to specify the value of $a$, designating which pair 
$k_{a-1}+k_a$ is joined.
In (\ref{hhh}), $a=4$, and the associated momenta and variables are 
$k_1,k_2,k_3+k_4,k_5,k_6$
with $z_1,z_2,z_4,z_5,z_6$ so
\begin{align}\label{h5}
{\h}^5_1&= s_{12}z_{24} -s_{234} z_{45} + s_{56} z_{45} - s_{61}z_5,
\cr {\h}^5_2&= s_{12} z_{24}z_5 -s_{345}z_{24}z_5 + s_{56} z_{45} - s_{61} z_4z_5,
\end{align}
which is the appropriately shifted version of
the non-gothic polynomials 
\begin{align}
h^5_1 &= s_{12}z_{23} - s_{23}z_{34} +s_{45}z_{34} -s_{51}z_4,
\cr h^5_2&= s_{12}z_{23}z_4
+ s_{34} z_{32}z_4 + s_{45}z_{34}-s_{51}z_4z_3.
\end{align}

 Now, more generally, suppose $a>2$ and consider $z_1\rightarrow\infty$,
  \begin{align}\label{Arel2}
  \left. h^N_{(z_a)}\right|_{z_{a-1}=z_a}
&=s_{a-1,a}   \prod_{b\ne 1,a-1,a} (z_b-z_a),\cr
\left. h^N_{(z_b)}\right|_{z_{a-1}=z_a}
&=(z_a-z_b) {\frak h}^{N-1}_{(z_b)}\,,
\hskip50truemm b\ne 1, a-2,a-1,a,a+1,\cr
\left. h^N_{(z_{a-2})}\right|_{z_{a-1}=z_a}
&=(z_a-z_{a-2}) {\frak h}^{N-1}_{(z_{a-2})}\cr
&\qquad+s_{a-1,a} (z_{a+1}-z_a)(z_{a-2}-z_a)  \prod_{b\ne 1,a-2,a-1,a,a+1} (z_b-z_{a-2}),\qquad a\ne3,\cr
\left. h^N_{(z_{a+1})}\right|_{z_{a-1}=z_a}
&=(z_a-z_{a+1}) {\frak h}^{N-1}_{(z_{a+1})}\cr
&\qquad+s_{a-1,a} (z_{a+1}-z_a)(z_{a-2}-z_a) \prod_{b\ne 1,a-2,a-1,a,a+1} (z_b-z_{a+1}),\qquad a\ne N,\cr
  \end{align}
  and we also have
  \be\label{A1}
  \left.h_1^N\right|_{z_{a-1}=z_a}= \h_{1}^{N-1}\,.
  \ee
In the particular case considered in subsection \ref{RR}, where we take $z_1\rightarrow\infty, z_2=1, z_3=x, z_4=y,z_N=0,$
 \begin{align}\label{Arel3}
  \left. h^N_{(y)}\right|_{x=y}
&=s_{34}   \prod_{b\ne 1,3,4} (z_b-y),\cr
\left. h^N_{(z_b)}\right|_{x=y}
&=(y-z_b) {\frak h}^{N-1}_{(z_b)}\,,
\hskip40truemm b\geq 6,\cr
\left. h^N_{(1)}\right|_{x=y}
&=(y-1) {\frak h}^{N-1}_{(1)}+s_{34} (z_{5}-y)(1-y)  \prod_{b\geq 6} (z_b-1),\cr
\left. h^N_{(z_{5})}\right|_{x=y}
&=(y-z_{5}) {\frak h}^{N-1}_{(z_{5})}+s_{34} (z_{5}-y)(1-y) \prod_{b\geq 6} (z_b-z_{5})\,.
  \end{align}
  
\section{An $\bf {N=8}$ Example}
 \label{Example}{}

As a further example, we consider the N=8 diagram $\widetilde M$ of Fig. 4a. After fixing M\"obius invariance, the CHY integral for the corresponding amplitude, $\A_8^{\widetilde M}$,
has three simple poles in the denominator, illustrating the growing complexity that is possible
as $N$ increases. At first glance, this would seem to make the evaluation of the integral using the global residue theorem more complicated. However, this complication is avoided by exchanging the 
scattering polynomials $h^N_a$ for the linear combinations $h^N_{(z_{a+3})}$, $2\leq a\leq N-3$, reducing the number of simple poles to just one, and thus
putting the integrand into a form where the global residue theorem 
can be applied simply. 
\begin{center}
\begin{tikzpicture}[scale=0.5]
\draw [thick] (0,0) --(1,1) -- (9,1) -- (10,0);
\draw [thick] (9,1) --(10,2);
\draw [thick] (0,2) --(1,1);
\draw [thick] (7,1) --(7,3)--(8,4);
\draw [thick] (7,3)--(6,4);
\draw [thick] (3,1) --(3,3)--(4,4);
\draw [thick] (3,3)--(2,4);
\node [left, blue] at (-0.4,0.3) {$\scriptstyle k_1$};
\node [left] at (0,-0.3) {$\scriptstyle z_1=\infty$};
\node [left, blue] at (-0.3,2.3) {$\scriptstyle  k_2$};
\node [left] at (0,1.7) {$\scriptstyle z_2=1$};
\node [above, blue] at (6,4.5) {$\scriptstyle  k_5$};
\node [above] at (6,3.9) {$\scriptstyle z_5$};
\node [above, blue] at (8,4.5) {$\scriptstyle  k_6$};
\node [above ] at (8,3.9) {$\scriptstyle z_6$};
\node [above, blue] at (2,4.5) {$\scriptstyle  k_3 $};
\node [above] at (2,3.9) {$\scriptstyle   z_3$};
\node [above, blue] at (4,4.5) {$\scriptstyle  k_4 $};
\node [above] at (4,3.9) {$\scriptstyle   z_4$};
\node [right, blue] at (10,2.3) {$\scriptstyle  k_7 $};
\node [right] at (10,1.7) {$\scriptstyle z_7$};
\node [right, blue] at (10.4,0.3) {$\scriptstyle  k_8$};
\node [right] at (10,-0.3) {$\scriptstyle  z_8=0$};
\node [below] at (5,-0.5) {$\widetilde M$};
\node [below] at (5,-2) {Fig. 4a};
\draw [thick] (16,0) --(17,1) -- (25,1) -- (26,0);?6
\draw [thick] (25,1) --(26,2);
\draw [thick] (16,2) --(17,1);
\draw [thick] (23,1) --(23,3)--(24,4);
\draw [thick] (23,3)--(22,4);
\draw [thick] (19,1) --(19,3);
\node [left, blue] at (15.6,0.3) {$\scriptstyle k_1$};
\node [left] at (16,-0.3) {$\scriptstyle z_1=\infty$};
\node [left, blue] at (15.7,2.3) {$\scriptstyle  k_2$};
\node [left] at (16,1.7) {$\scriptstyle z_2=1$};
\node [above, blue] at (22,4.5) {$\scriptstyle  k_5$};
\node [above] at (22,3.9) {$\scriptstyle z_5$};
\node [above, blue] at (24,4.5) {$\scriptstyle  k_6$};
\node [above ] at (24,3.9) {$\scriptstyle z_6$};
\node [right, blue] at (26,2.3) {$\scriptstyle  k_7 $};
\node [right] at (26,1.7) {$\scriptstyle z_7$};
\node [right, blue] at (26.4,0.3) {$\scriptstyle  k_8$};
\node [right] at (26,-0.3) {$\scriptstyle  z_8=0$};
\node [above, blue] at (19,3.5) {$\scriptstyle  k_3+k_4$};
\node [above] at (19,2.9) {$\scriptstyle  z_4$};
\node [below] at (21,-0.5) {$\widetilde M'$};
\node [below] at (21,-2) {Fig. 4b};
\end{tikzpicture}
\end{center}

Writing $z_{ab}=z_a-z_b,$
\begin{align}
\A_8^{\widetilde M}&=\oint {dz_3dz_4dz_5dz_6dz_7\over h^8_1h^8_2h^8_3h^8_4h^8_5}
{z_4^2z_6z_{24}z_{26}z_{27}z_{35}z_{36}z_{37}z_{46}z_{57}\over z_{34}z_{56}z_7}\cr
&=\oint {dz_3dz_4dz_5dz_6dz_7\over h^8_1 h^8_{(z_5)}h^8_{(z_6)}h^8_{(z_7)}h^8_5}
{z_4^2z_5z^2_6z_{24}z_{26}z_{27}z_{35}z_{36}z_{37}z_{46}z^2_{57}z_{67}
\over z_{34}}\cr
&=-\oint {dz_3dz_4dz_5dz_6dz_7\over h^8_1 z_{34} h^8_{(z_6)}h^8_{(z_7)}h^8_5}
{z_4^2z_5z_6^2z_{24}z_{26}z_{27}z_{45}z^2_{46}z_{47}z_{57}^2 z_{67}\over
h^8_{(z_5)}}\cr
&={1\over s_{34}}\oint {dz_4dz_5dz_6dz_7 
\over \h^7_1\h^7_{(z_6)} \h^7_{(z_7)}\h^7_4 }
{z_4z_6^2z_{26}z_{27}z_{46}z_{57}z_{67}\over z_{56}}\cr
&=-{1\over s_{34}}\oint {dz_4dz_5dz_6dz_7\over \h^7_1\h^7_2\h^7_3\h^7_4 }
{z_4z_6z_{26}z_{27}z_{46}z_{57}\over z_{56}z_7} =-{1\over s_{34}}\A_7^{\widetilde M'}\,,
\end{align}
where $\A_7^{\widetilde M'}$ is the amplitude associated with Fig. 4b, 
since, when $z_{34}=0$,
\begin{align}
h_1^8&=\h_1^7,\quad h^8_{(z_5)} = s_{34}z_{35}z_{23}z_{65}z_{75}z_5,
\quad h^8_{(z_6)} = z_{36}\h^7_{(z_6)},\quad
h^8_{(z_7)} = z_{37}\h^7_{(z_7)},\quad
h^8_5 = z_4 \h^7_4.
\nonumber\end{align}

\end{appendices}

\vskip20pt
\singlespacing

\vfil\eject

\providecommand{\bysame}{\leavevmode\hbox to3em{\hrulefill}\thinspace}
\providecommand{\MR}{\relax\ifhmode\unskip\space\fi MR }
\providecommand{\MRhref}[2]
{
}
\providecommand{\href}[2]{#2}

\end{document}